\overfullrule=0pt
\input harvmac

\lref\gutt{
  P.~A.~Grassi and S.~Guttenberg,
``On Projections to the Pure Spinor Space,''
JHEP {\bf 1112}, 089 (2011).
[arXiv:1109.2848 [hep-th]]\semi
S.~Guttenberg,
``A Projection to the Pure Spinor Space,''
[arXiv:1202.0335 [hep-th]].}

\lref\Aisaka{
  Y.~Aisaka and Y.~Kazama,
  ``Origin of pure spinor superstring,''
JHEP {\bf 0505}, 046 (2005).
[hep-th/0502208].
}

\lref\BerkovitsGH{
  N.~Berkovits,
  ``Pure spinors, twistors, and emergent supersymmetry,''
JHEP {\bf 1212}, 006 (2012).
[arXiv:1105.1147 [hep-th]].
}

\lref\Matone{
  M.~Matone, L.~Mazzucato, I.~Oda, D.~Sorokin and M.~Tonin,
  ``The Superembedding origin of the Berkovits pure spinor covariant quantization of superstrings,''
Nucl.\ Phys.\ B {\bf 639}, 182 (2002).
[hep-th/0206104].
}

\lref\Bnon{
  N.~Berkovits,
  ``Pure spinor formalism as an N=2 topological string,''
JHEP {\bf 0510}, 089 (2005).
[hep-th/0509120].
}

\lref\BerkovitsPLA{
  N.~Berkovits,
  ``Dynamical twisting and the b ghost in the pure spinor formalism,''
JHEP {\bf 1306}, 091 (2013).
[arXiv:1305.0693 [hep-th]].
}

\lref\Bpure{
  N.~Berkovits,
  ``Super Poincare covariant quantization of the superstring,''
JHEP {\bf 0004}, 018 (2000).
[hep-th/0001035].
}

\lref\Donagi{
  R.~Donagi and E.~Witten,
``Supermoduli Space Is Not Projected,''
[arXiv:1304.7798 [hep-th]]\semi
  E.~Witten,
  ``More On Superstring Perturbation Theory,''
[arXiv:1304.2832 [hep-th]]\semi
  E.~Witten,
  ``Superstring Perturbation Theory Revisited,''
[arXiv:1209.5461 [hep-th]].
}

\lref\explaining {
N.~Berkovits,
``Explaining the Pure Spinor Formalism for the Superstring,''
JHEP {\bf 0801}, 065 (2008).
[arXiv:0712.0324 [hep-th]].
}

\lref\FriedanGE{
  D.~Friedan, E.~J.~Martinec and S.~H.~Shenker,
  ``Conformal Invariance, Supersymmetry and String Theory,''
Nucl.\ Phys.\ B {\bf 271}, 93 (1986).
}

\lref\Baulieua{
 L.~Baulieu,
 ``Transmutation of pure 2-D supergravity into topological 2-D gravity and other conformal theories,''
Phys.\ Lett.\ B {\bf 288}, 59 (1992).
[hep-th/9206019]\semi
  L.~Baulieu, M.~B.~Green and E.~Rabinovici,
  ``A Unifying topological action for heterotic and type II superstring theories,''
Phys.\ Lett.\ B {\bf 386}, 91 (1996).
[hep-th/9606080]\semi
 L.~Baulieu, M.~B.~Green and E.~Rabinovici,
  ``Superstrings from theories with $N>1$ world sheet supersymmetry,''
Nucl.\ Phys.\ B {\bf 498}, 119 (1997).
[hep-th/9611136]\semi
  L.~Baulieu and N.~Ohta,
  ``World sheets with extended supersymmetry,''
Phys.\ Lett.\ B {\bf 391}, 295 (1997).
[hep-th/9609207].
}

\lref\osv{
  O.~Chandia,
  ``The b Ghost of the Pure Spinor Formalism is Nilpotent,''
Phys.\ Lett.\ B {\bf 695}, 312 (2011).
[arXiv:1008.1778 [hep-th]].
}

\lref\WittenMB{
  E.~Witten,
  ``D = 10 Superstring Theory,''
In *Philadelphia 1983, Proceedings, Grand Unification*, 395-408.
}

\lref\rennan{
  R.~L.~Jusinskas,
  ``Nilpotency of the b ghost in the non minimal pure spinor formalism,''
[arXiv:1303.3966 [hep-th]].
}

\lref\BeisertJR{
  N.~Beisert, C.~Ahn, L.~F.~Alday, Z.~Bajnok, J.~M.~Drummond, L.~Freyhult, N.~Gromov and R.~A.~Janik {\it et al.},
  ``Review of AdS/CFT Integrability: An Overview,''
Lett.\ Math.\ Phys.\  {\bf 99}, 3 (2012).
[arXiv:1012.3982 [hep-th]].
}

\lref\Grassia{
  P.~A.~Grassi, G.~Policastro, M.~Porrati and P.~Van Nieuwenhuizen,
  ``Covariant quantization of superstrings without pure spinor constraints,''
JHEP {\bf 0210}, 054 (2002).
[hep-th/0112162]\semi
  P.~A.~Grassi, G.~Policastro and P.~van Nieuwenhuizen,
  ``On the BRST cohomology of superstrings with / without pure spinors,''
Adv.\ Theor.\ Math.\ Phys.\  {\bf 7}, 499 (2003).
[hep-th/0206216].
}

\lref\Matone{
  M.~Matone, L.~Mazzucato, I.~Oda, D.~Sorokin and M.~Tonin,
  ``The Superembedding origin of the Berkovits pure spinor covariant quantization of superstrings,''
Nucl.\ Phys.\ B {\bf 639}, 182 (2002).
[hep-th/0206104].
}

\lref\Tonin{
  M.~Tonin,
  ``World sheet supersymmetric formulations of Green-Schwarz superstrings,''
Phys.\ Lett.\ B {\bf 266}, 312 (1991).
}

\lref\Sorokin{
  D.~P.~Sorokin,
  ``Superbranes and superembeddings,''
Phys.\ Rept.\  {\bf 329}, 1 (2000).
[hep-th/9906142].
}

\lref\Aisakaa{
  Y.~Aisaka and Y.~Kazama,
  ``A new first class algebra, homological perturbation and extension of pure spinor formalism for superstring,''
JHEP {\bf 0302}, 017 (2003).
[hep-th/0212316]\semi
  Y.~Aisaka and Y.~Kazama,
  ``Operator mapping between RNS and extended pure spinor formalisms for superstring,''
JHEP {\bf 0308}, 047 (2003).
[hep-th/0305221].
}

\lref\BerkovitsWR{
  N.~Berkovits,
  ``Covariant quantization of the Green-Schwarz superstring in a Calabi-Yau background,''
Nucl.\ Phys.\ B {\bf 431}, 258 (1994).
[hep-th/9404162].
}

\lref\BerkovitsUS{
  N.~Berkovits,
  ``Relating the RNS and pure spinor formalisms for the superstring,''
JHEP {\bf 0108}, 026 (2001).
[hep-th/0104247].
}

\lref\BerkovitsU{
  N.~Berkovits,
  ``Quantization of the superstring with manifest U(5) superPoincare invariance,''
Phys.\ Lett.\ B {\bf 457}, 94 (1999).
[hep-th/9902099].
}

\lref\BerkovitsVI{
  N.~Berkovits and N.~Nekrasov,
  ``Multiloop superstring amplitudes from non-minimal pure spinor formalism,''
JHEP {\bf 0612}, 029 (2006).
[hep-th/0609012].
}

\lref\SiegelT{
  W.~Siegel,
  ``Superfields in Higher Dimensional Space-time,''
Phys.\ Lett.\ B {\bf 80}, 220 (1979).
}

\lref\WittenT{
  E.~Witten,
  ``Twistor - Like Transform in Ten-Dimensions,''
Nucl.\ Phys.\ B {\bf 266}, 245 (1986).
}

\lref\BerkovitsBT{
  N.~Berkovits,
  ``Pure spinor formalism as an N=2 topological string,''
JHEP {\bf 0510}, 089 (2005).
[hep-th/0509120].
}

\lref\BerkovitsNek{
  N.~Berkovits and N.~Nekrasov,
  ``Multiloop superstring amplitudes from non-minimal pure spinor formalism,''
JHEP {\bf 0612}, 029 (2006).
[hep-th/0609012].
}

\lref\siegelclassical{
  W.~Siegel,
  ``Classical Superstring Mechanics,''
Nucl.\ Phys.\ B {\bf 263}, 93 (1986).
}

\lref\Baulieuf{
  L.~Baulieu,
  ``SU(5)-invariant decomposition of ten-dimensional Yang-Mills supersymmetry,''
Phys.\ Lett.\ B {\bf 698}, 63 (2011).
[arXiv:1009.3893 [hep-th]].
}

\lref\SiegelYD{
  W.~Siegel,
[arXiv:1005.2317 [hep-th]].
}

\def\bar{\overline}

\def\a{{\alpha}}

\def\l{{\lambda}}
\def\lb{{\overline\lambda}}
\def\wb{{\overline w}}
\def\wtb{{\widetilde{\overline w}}}

\def\lb{{\overline\lambda}}
\def\b{{\beta}}
\def\bh{{\widehat\beta}}

\def\g{{\gamma}}
\def\G{{\Gamma}}
\def\Gb{{\bar\Gamma}}
\def\gh{{\widehat\gamma}}

\def\d{{\delta}}
\def\e{{\epsilon}}
\def\s{{\sigma}}

\def\L{{\Lambda}}
\def\O{{\Omega}}
\def\half{{1\over 2}}
\def\p{{\partial}}

\def\t{{\theta}}

\def\lb{{\bar{\lambda}}}

\Title{\vbox{\baselineskip12pt
\hbox{ICTP-SAIFR/2013-014 }}}
{{\vbox{\centerline{Covariant Map Between Ramond-Neveu-Schwarz}
\smallskip
\centerline{and Pure Spinor Formalisms for the Superstring}}} }
\bigskip\centerline{Nathan Berkovits\foot{e-mail: nberkovi@ift.unesp.br}}
\bigskip
\centerline{\it ICTP South American Institute for Fundamental Research}
\centerline{\it Instituto de F\'\i sica Te\'orica, UNESP - Univ. 
Estadual Paulista }
\centerline{\it Rua Dr. Bento T. Ferraz 271, 01140-070, S\~ao Paulo, SP, Brasil}
\bigskip

\vskip .3in

A covariant map between the Ramond-Neveu-Schwarz (RNS) and pure spinor formalisms for the superstring is found which transforms the RNS and pure spinor BRST operators into each other. The key ingredient is a dynamical twisting of the ten spin-half RNS fermions
into five spin-one and five spin-zero fermions using bosonic pure spinors that
parameterize an $SO(10)/U(5)$ coset. The map relates massless vertex operators in the two formalisms, and gives a new description of Ramond states which does
not require spin fields. An argument is proposed for relating the amplitude prescriptions in the two formalisms.
\vskip .3in

\Date {December 2013}

\newsec{Introduction}

Although the Ramond-Neveu-Schwarz (RNS) formalism for the superstring has an
elegant worldsheet description as an N=1 superconformal field theory, its spacetime
description is complicated. Vertex operators for Ramond states require spin fields and it is unknown how to describe the RNS formalism in Ramond-Ramond
backgrounds. Furthermore, the RNS scattering amplitude prescription requires summing
over spin structures to project out states in the $GSO(-)$ sector, and cancellations implied by 
spacetime supersymmetry are far from manifest.

On the other hand, the pure spinor formalism for the superstring has an elegant spacetime
description in which vertex operators are expressed in d=10 superspace and spacetime
supersymmetry is manifest. However, its worldsheet description is mysterious and the pure
spinor BRST operator has not yet been derived from gauge-fixing a worldsheet
reparamaterization-invariant action. 

Constructing a map between these two superstring formalisms is an obvious way to better understand both formalisms. In light-cone gauge, the pure spinor formalism is equivalent to the light-cone Green-Schwarz (GS) formalism which was mapped in \WittenMB\ to light-cone RNS. This map manifestly preserves an SU(4) subgroup
of the SO(8) light-cone symmetry and transforms the eight light-cone RNS vector variables
$\psi^j$ into the eight light-cone GS spinor variables $\t^a$. Splitting the SO(8) vector
$\psi^j$ and SO(8) spinor $\t^a$ as $(\psi^J,\psi_J)$ and $(\t^J, \t_J)$ where $J=1$
to 4 is an SU(4) index, the map of \WittenMB\ is obtained by bosonizing
\eqn\bosoniz{\psi^J = e^{i\s_J}, \quad \psi_J = e^{-i\s_J}}
and writing $(\t^J, \t_J)$ as the spin fields
$$\t^J = e^{i\s_J - {i\over 2} (\s_1 + \s_2 +\s_3 + \s_4)}, \quad 
\t_J = e^{-i\s_J + {i\over 2} (\s_1 + \s_2 +\s_3 + \s_4)}.$$
Note that all $(\psi^J, \psi_J)$ and $(\t^J, \t_J)$ variables carry conformal weight $\half$.

To find a covariant version of this map, the first step is to enlarge the SU(4) 
symmetry of \bosoniz\ to an SU(5) subgroup of the (Wick-rotated) SO(10) Lorentz group.
In the ``U(5) hybrid formalism" of \BerkovitsU, this was done by splitting the RNS SO(10)
vector $\psi^m$ for $m=0$ to 9 into $(\psi^A, \psi_A)$ where $A=1$ to 5 is an SU(5) index, bosonizing
as 
\eqn\ddp{\psi^A = e^{i\s_A}, \quad \psi_A = e^{-i\s_A},}
and constructing 5 of the 16 components of the SO(10) spinor $\t^\a$ and its
conjugate momentum $p_\a$ as the spin fields
\eqn\dbt{\t^A = e^{i\s_A - {i\over 2}(\s_1 +\s_2+\s_3+\s_4+\s_5)} e^{\half\phi}, \quad
p_A = e^{-i\s_A + {i\over 2}(\s_1 +\s_2+\s_3+\s_4+\s_5)} e^{-\half\phi} }
where $\phi$ comes from the Friedan-Martinec-Shenker bosonization \FriedanGE\ of the
$(\b,\g)$ ghosts as $\b\g = \p\phi$. Note that $\t^A$ carries conformal weight zero as in the
covariant GS formalism and its conjugate momentum $p_A$
carries conformal
weight $+1$. However, the other 11 components of the SO(10) spinors
$\t^\a$ and $p_\a$ were absent in this U(5) hybrid formalism so SO(10)
covariance was not manifest.

In the pure spinor formalism for the superstring, all 16 components of $\t^\a$ and $p_\a$
are present as well as the 11 independent components of a bosonic pure spinor $\l^\a$
satisfying $\l\g^m \l =0$ and its conjugate momentum $w_\a$. A map was proposed in \BerkovitsUS\ between the RNS and pure spinor
formalism which combined the U(5) hybrid formalism with a
``topological'' sector containing $(\l^\a, w_\a)$ and the 11 remaining components of
$(\t^\a, p_\a)$. However, because of the complicated bosonization formula of \dbt\ used
in the hybrid formalism,
the map was not manifestly SO(10) covariant. Although there exists a relation at the classical
level between the hybrid formalism and the manifestly covariant ``superembedding'' formalisms \Tonin\Matone\Sorokin, this relation has not yet been understood at the quantum level. 

In \explaining, a new approach to relating the RNS and pure spinor variables was proposed in
which bosonization of the RNS ghost and matter fields is unnecessary. In this approach,
one simply rescales the U(5) components $(\psi^A,\psi_A)$ of $\psi^m$ in opposite
directions using the $\g$ ghost as
\eqn\twg{\G^A = \g \psi^A, \quad \Gb_A = {1\over \g} \psi_A,}
where $\G^A$ and $\Gb_A$ are GSO-even fermions of conformal weight 0 and 1.
This ``twisting'' by the $\g$ ghost was earlier used in the Calabi-Yau fermions of the
d=4 hybrid formalism \BerkovitsWR\ and also appeared in the topological twisting papers of
Baulieu et al \Baulieua. Although \twg\ breaks SO(10) covariance
to U(5), one can recover the full SO(10) covariance by using the pure spinor $\l^\a$
and its complex conjugate $\lb_\a$ to ``dynamically'' choose the U(5) direction of the twisting so that
\eqn\twgn{\G^m = \g {{\l\g^m\g^n\lb}\over {2(\l\lb)} }\psi_n,
 \quad \Gb_m = {1\over \g}{{\l\g_n\g_m\lb}\over {2(\l\lb)}}  \psi^n.}
Note that $\G^m$ and $\Gb_m$ each have five independent components since they satisfy
$\G^m (\g_m\l)_\a = \Gb_m (\g^m\lb)^\a =0$, and were related in \explaining\
to five components of $\t^\a$ and $p_\a$. 

Although both the bosonization formulas of \dbt\ and the twisting formulas of \twg\ and \twgn\
map RNS spin-half fermions into GS-like spin-zero and spin-one fermions, the relation of the
two maps is unclear. As stressed by Witten \ref\wittenp{E. Witten, private communication.}, 
bosonization formulas such as \dbt\ can be ill-defined at higher genus, but this does not seem
to be a problem for the twisting formulas of \twg\ and \twgn.
The rigid twisting of \twg\ is related to holomorphic d=5 super-Yang-Mills
\ref\nekr{N. Nekrasov, KITP 2009 Lecture, ``Pure spinors, beta-gammas, super-Yang-Mills and Chern-Simons, Part 2'', http://online.kitp.ucsb.edu/online/strings09/nekrasov2/.}\Baulieuf, and  it was conjectured by Nekrasov in \nekr\ that the dynamical twisting of \twgn\ replaces the topogical spectrum of holomorphic d=5 super-Yang-Mills with the
d=10 superstring spectrum. Evidence for Nekrasov's conjecture was obtained recently in
\BerkovitsPLA\ where the ``dynamical twisting'' of \twgn\ was shown to simplify the expression for the composite $b$ ghost in the pure spinor formalism. And in this paper, Nekrasov's conjecture
will be confirmed by showing that it maps the RNS and pure spinor BRST operators into each other.

To use the dynamical twisting procedure of \twgn\ to provide a covariant map between
the RNS and pure spinor BRST operators, the first step will be to add to the usual RNS
variables a topological set of ``non-minimal'' fermionic and bosonic spacetime spinor
variables $(\t^\a, p_\a)$ and $(\L^\a, \O_\a)$ of conformal weight $(0,1)$. The BRST operator
in this ``non-minimal'' RNS formalism will be defined as 
\eqn\nonb{Q= Q_{RNS} + 
\int dz (\L^\a p_\a)}
 so that the cohomology of physical states is unchanged.
After the similarity transformation $Q \to e^{-R} Q e^R$ where 
\eqn\simsim{R= {1\over {2\g}}  (\L\g_m\t)\psi^m,}
the non-minimal BRST operator of \nonb\ can be surprisingly written in manifestly
spacetime supersymmetric form where $(x^m,\t^\a)$ transform as d=10 superspace
variables. Furthermore, despite the presence of $1\over \g$ in \simsim, vertex operators
in the $GSO(+)$ sector can be written in d=10 superspace and do not contain any
inverse powers of $\g$. So Ramond states in this non-minimal RNS formalism do not
require spin fields or bosonization and one can easily describe the formalism in curved
Ramond-Ramond backgrounds.

To perform dynamical twisting as in \twgn, one decomposes the unconstrained bosonic
spinor $\L^\a$ into pure spinors $\l^\a$ and $\lb_\a$ by defining \gutt
\eqn\uncl{\L^\a = \l^\a + {1\over{2(\l\lb)}} u^m (\g_m\lb)^\a}
where $u^m$ is a bosonic vector with only five independent components because of the
gauge invariance $\d u^m = (\e\g^m\lb)$. The definition of \uncl\
for the unconstrained $\L^\a$ might be useful for understanding the relation with ``extended''
versions of the pure spinor formalism such as \Grassia\Aisakaa\  in which the spinor ghosts were unconstrained. After defining $\G^m$ and $\Gb_m$ as in \twgn, the RNS
$\g$ ghost only appears in even powers so it is convenient to define a new
ghost variable $\gh \equiv (\g)^2$. Since $\gh$ and its conjugate momentum $\bh$ carry
conformal weight $-1$ and $+2$, the contribution of $(\G^m, \Gb_m)$ and $(\bh,\gh)$ to the conformal anomaly is $-10 + 26$ which is equal to the $+5 + 11$ contribution of the
original $\psi^m$ and $(\b,\g)$ variables.

Expressing the non-minimal RNS BRST operator in terms of $(\G^m,\Gb_m)$ and
$(\bh,\gh)$ and decomposing $\L^\a$ as in \uncl, one finds after a similarity transformation that 
\eqn\purebr{Q = \int dz (\l^\a d_\a + \wb^\a r_\a + u^m \Gb_m + \gh b)}
where  $\int dz (\l^\a d_\a)$ is the pure spinor BRST operator.
So after adding non-minimal spinor variables to the RNS formalism, dynamically twisting
as in \twgn, and performing various similarity transformations, the RNS BRST operator
is covariantly mapped into the pure spinor BRST operator plus a set of ``topological'' variables
which decouple from the cohomology.
Furthermore, the non-minimal RNS formalism containing both the RNS $\psi^m$ variables and
the GS $\t^\a$ variables provides a natural bridge between the RNS and pure spinor
formalisms which resembles the ``superembedding'' formalisms reviewed in \Sorokin. 
Vertex operators in the $GSO(+)$ sector can be expressed in d=10 superspace
using the non-minimal RNS formalism, and in the gauge $u^m =\G^m=0$, they reduce to the
usual pure spinor vertex operators. And in the gauge $\L^\a =\t^\a =0$, the vertex operators for
bosons in the non-minimal RNS formalism reduce to the usual Neveu-Schwarz vertex operators of the RNS formalism in the zero picture.

The covariant map can also be used to relate the scattering amplitude prescriptions in
the RNS and pure spinor formalisms. Both of these formalisms contain chiral bosons,
and functional integration over chiral bosons is divergent because of their non-compact
zero modes. These divergences are cancelled by zeros coming from functional integration
over the zero modes of chiral fermions, and a convenient BRST-invariant method for
regularizing the divergence is to insert a picture-changing operator for each chiral boson
zero mode. Since the dynamical twisting procedure changes the $(\b,\g)$ chiral bosons of the
RNS formalism to $(\bh,\gh)$ chiral bosons which carry different conformal weight, the number and type of
picture-changing operators inserted in the RNS and pure spinor formalism are different. Nevertheless, assuming that the dynamical twisting procedure is a consistent field redefinition at the quantum level, one expects that the different RNS and pure spinor prescriptions for regularizing the chiral boson zero modes should lead to the same scattering amplitude.

In section 2 of this paper, the non-minimal RNS formalism is constructed and the dynamical
twisting procedure is defined which covariantly maps the RNS BRST operator into
the pure spinor BRST operator. In section 3, the massless vertex operators in the
non-minimal RNS formalism are constructed and shown to form a bridge between the
RNS and pure spinor vertex operators. And in section 4, the RNS and pure spinor
scattering amplitude prescriptions are related to each other through the dynamical twisting procedure.

\newsec{Covariant Map}

\subsec{Non-minimal RNS formalism}

The usual RNS worldsheet action, stress tensor, and BRST operator are
\eqn\rnsa{S_{RNS} = \int d^2 z (\half \p x^m \bar\p x_m +\half \psi^m \bar\p \psi_m
+\b\bar\p \g + b\bar\p c),}
\eqn\rnsb{ T_{RNS} = -\half \p x^m \p x_m - \half\psi^m \p \psi_m
-\b\p \g -\half \p(\b\g) - b\p c - \p(b c),}
\eqn\rnsc{Q_{RNS} = \int dz ( c T_{RNS} + \g \psi^m \p x_m + \g^2 b - b c \p c )}
where the right-moving variables $(\bar\psi^m, \bar\beta,\bar\g,\bar b,\bar c)$
will be ignored throughout this paper. The free field OPE's of the left-moving
RNS variables of \rnsa\ are
\eqn\fone{\p x^m (z) \p x^n (0) \to - z^{-2} \eta^{mn}, \quad
\psi^m(z) \psi^n(0) \to z^{-1} \eta^{mn},}
$$ \g(z)\b(0) \to z^{-1}, \quad c(z) b(0) \to z^{-1}.$$
Although only the open superstring will be discussed
in this paper, all results can be easily generalized
to the closed superstring by taking the ''left-right
product'' of two open superstrings.

To relate \rnsa, \rnsb\ and \rnsc\ to the pure spinor worldsheet action, stress tensor, and BRST operator,
the first step is to add a non-minimal set of  fermionic spacetime spinor variables $(\t^\a, p_\a)$ of conformal weight $(0,1)$ and bosonic unconstrained spacetime spinor variables $(\L^\a, \O_\a)$ of conformal weight $(0,1)$ so that 
\eqn\nrnsa{S = S_{RNS} + \int d^2 z (p_\a \bar\p \t^\a + \O_\a \bar\p \L^\a), }
\eqn\nrnsb{ T = T_{RNS} - p_\a \p\t^\a - \O_\a \p\L^\a,}
\eqn\nrnsc{Q = Q_{RNS} + \int dz (\L^\a p_\a),}
with the free field OPE's
\eqn\ftwo{ \L^\a(z)\O_\b(0) \to z^{-1} \d^\a_\b, \quad \t^\a(z) p_\b(0) \to z^{-1} \d^\a_\b.}
Using the usual quartet argument, the BRST cohomology is unchanged. Performing
the similarity transformation ${\cal O} \to e^{-R} {\cal O} e^R$ on all operators ${\cal O}$
where 
\eqn\simone{R = \int dz ~c \O_\a \p\t^\a,}
the BRST operator of \nrnsc\ is transformed into the more
conventional form
\eqn\nrnsd{Q = \int dz (\L^\a p_\a + c T +   \g \psi^m \p x_m + \g^2 (b+\O_\a\p\t^\a) - b c \p c )}
where $T$ is defined in \nrnsb.

Since the worldsheet variables include the $(\t^\a, p_\a)$ variables of d=10 superspace,
one can construct the operators\siegelclassical\ 
\eqn\susyg{q_\a = \int dz ( p_\a + \half (\p x^m +{1\over{12}}\t\g^m \p\t)(\g_m\t)_\a )}
which generate the d=10 spacetime supersymmetry algebra $\{q_\a,q_\b\} = \g^m_{\a\b}
\int dz ~\p x_m$. Although $q_\a$ does not anticommute with the BRST operator of \nrnsd,
one can perform the further similarity transformation
${\cal O} \to e^{-R'} {\cal O} e^{R'}$ where
\eqn\simtwo{R' = \int dz {1\over {2\g}}(\L\g^m \t)\psi_m,}
under which the BRST operator of \nrnsd\ transforms into the manifestly spacetime
supersymmetric operator
\eqn\rnse{Q = \int dz (\L^\a d_\a +{1\over {2\g}} (\L\g^m\L)\psi_m + c T +   \g \psi^m \Pi_m + \g^2 (b+\O_\a\p\t^\a) - b c \p c )}
where 
\eqn\defd{d_\a = p_\a -\half (\p x^m +{1\over 4} \t\g^m \p\t)(\g_m\t)_\a , \quad
\Pi^m = \p x^m +\half\t\g^m \p\t}
are the usual spacetime supersymmetric operators \siegelclassical\ for fermionic and bosonic momenta.
Note that $T$ of \nrnsb\ can be written in the manifestly spacetime supersymmetric form 
\eqn\tsusy{T = -\half \Pi^m \Pi_m - d_\a \p\t^\a -  \O_\a \p\L^\a - \half\psi^m \p \psi_m
-\b\p \g -\half \p(\b\g) - b\p c - \p(b c).}

So after adding the non-minimal spinor variables and performing the similarity
transformation of \simtwo, the non-minimal RNS BRST operator and stress tensor of \rnse\ and
\tsusy\ are manifestly invariant under the spacetime supersymmetry generated by \susyg.
But because of the inverse power of $\g$ in the similarity transformation of \simtwo\ and in
the term ${1\over {2\g}} (\L\g^m\L)\psi_m$ in the BRST current, the Hilbert space of states in the non-minimal
RNS formalism
is no longer the usual ``small'' Hilbert space of the RNS formalism in which all states are polynomials in $\b$ and $\g$. Nevertheless, it will be shown in section 3
that all states in the $GSO(+)$ sector of the
non-minimal RNS formalism can be described in the ``small'' Hilbert space and that spacetime supersymmetry acts covariantly on these states. Furthermore, it will now be shown that after twisting the ten spin-half
variables $\psi^m$ into five spin-zero and five spin-one variables, the inverse powers of
$\g$ can be eliminated from the BRST operator and the resulting
twisted version of the non-minimal RNS formalism is the pure spinor formalism.

\subsec{ Dynamical twisting}

To covariantly twist the ten $\psi^m$ spin-half variables into five spin-zero and five
spin-one variables, one needs to construct pure spinor variables $(\l^\a, \lb_\a)$ satisfying
the constraints
\eqn\purec{\l\g^m \l =0, \quad \lb\g^m\lb =0,}
whose 11 complex components (in Wick-rotated Euclidean space) parameterize the
complex space
${{SO(10)}\over {U(5)}}\times C$. In terms of the unconstrained spinor $\L^\a$, $\l^\a$ and
$\lb_\a$ will be defined as \gutt
\eqn\defL{\L^\a = \l^\a + {1\over{2(\l\lb)}} u^m (\g_m\lb)^\a}
where $u^m$ is a bosonic vector defined up to the gauge transformation
\eqn\gaugeu{\d u^m = \e_\a (\g^m \lb)^\a.}
Note that $\L^\a$ in \defL\ is unchanged under the gauge transformation
\eqn\gaugexi{\d \lb_\a = \xi_\a, \quad \d u^m =-{1\over {2(\l\lb)}} (\l\g^m \g^n \xi) u_n, \quad
\d \l^\a = {1\over {16(\l\lb)^2}}(\lb \g^{mn} \g^p \xi)(\g_{mn}\l)^\a u_p,}
where $\xi_\a$ is any spinor satisfying $\lb\g^m\xi =0$. The gauge transformations of
\gaugeu\ and \gaugexi\ can be used to gauge-fix all 11 components of $\lb_\a$ and
5 components of $u^m$, and the remaining 16 components of $u^m$ and $\l^\a$
are determined by $\L^\a$.

Defining $w_\a$ and $v_m$ to be the conjugate momenta to $\l^\a$ and $u_m$, one
finds that
\eqn\deffO{\O_\a = {1\over{4(\l\lb)}}[(\lb\g^{mn})_\a N_{mn} + \lb_\a (J + 4 u_m v^m) ] +
(\l\g^m)_\a v_m}
satisfies the desired OPE $\L^\a(z) \O_\b(0)\to z^{-1}\d^\a_\b$ where
$$N_{mn}= \half (\l\g_{mn} w), \quad J = -\l^\a w_\a.$$
Note that the gauge invariance of \gaugeu\ implies that $v^m$ is constrained to satisfy
\eqn\constv{ v^m (\g_m\lb)^\a =0.} 
To include the new variables in the formalism, first add $(\lb_\a, \widehat{\wb}^\a)$ and $(r_\a, s^\a)$ to \nrnsa, \nrnsb\ and \rnse\ as
\eqn\acttwoa{S = S_{RNS} + \int d^2 z (p_\a \bar\p\t^\a +  \O_\a \bar\p \L^\a + \widehat{\wb}^\a \bar\p \lb_\a + s^\a \bar\p r_\a ),}
\eqn\stresstwoa{T = T_{RNS} -p_\a \p\t^\a - \O_\a \p \L^\a - \widehat{\wb}^\a \p \lb_\a - s^\a \p r_\a, }
\eqn\brsttwoa{Q = \int dz (\L^\a d_\a  +{1\over {2\g}} (\L\g^m\L)\psi_m + c T +   \g \psi^m \Pi_m + \g^2 (b+\O_\a\p\t^\a) - b c \p c )}
$$+\int dz ( \widehat{\wb}^\a r_\a + \g^2 s^\a \p \lb_\a )$$
where $\widehat{\wb}^\a$ has no singular OPE's with $\L^\a$ or $\O_\a$, and
$r_\a$ is the fermionic ghost coming from the gauge parameter $\xi_\a$ of \gaugexi\ which
is constrained to satisfy 
\eqn\constr{r\g^m \lb =0.}
One can then plug into \acttwoa\ and \stresstwoa\ the expressions of \defL\ and
\deffO\ for $\L^\a$ and $\O_\a$ to find that
\eqn\acttwob{S = S_{RNS} + \int d^2 z (p_\a \bar\p\t^\a +  w_\a \bar\p \l^\a + v_m \bar\p u^m + \wb^\a \bar\p \lb_\a + s^\a \bar\p r_\a ),}
\eqn\stresstwob{T = T_{RNS} - p_\a \p\t^\a - w_\a \p \l^\a - v_m \p u^m - \wb^\a \p \lb_\a - s^\a \p r_\a, }
where $\wb^\a$ has no singular OPE's with $\l^\a$ or $u_m$ and is defined by
\eqn\defwb{\wb^\a = \widehat{\wb}^\a + {1\over{4(\l\lb)^2}} u^n
[ (\l\g_m \g_n\lb) (\g^m w)^\a - 2 (w\g_n\lb)\l^\a - 2 (\l\lb) v^m (\g_m\g_n\l)^\a ].}
Finally, 
the BRST operator can be expressed in terms of the pure spinor variables and
their conjugate momenta by plugging into \brsttwoa\
the expressions of \defL, \deffO, and \defwb\ for $\L^\a$, $\O_\a$ and $\widehat{\wb}^\a$ to obtain
\eqn\brstfinal{Q = 
 \int dz (\l^\a d_\a +\wb^\a r_\a +\g \psi^m \Pi_m + c T - b c \p c}
 $$+ \gh [b+ s^\a \p\lb_\a +
 {1\over{4(\l\lb)}}((\lb \g^{mn}\p\t)N_{mn} + (\lb \p\t) (J+4 u^n v_n)) + (\l\g^m\p\t) v_m ]$$
 $$+ u_m [ {1\over{2\g(\l\lb)}}(\lb\g^m \g^n \l)\psi_n +{1\over{2(\l\lb)}}\lb\g^m d - {{(\lb\g^{m} \g^{np} r)}\over{8(\l\lb)^2}} N_{np}
 + {{(\l\g^m\g^n r)}\over{2(\l\lb)}} v_n ]).$$

Using the pure spinor variables $(\l^\a,\lb_\a)$ to covariantly choose the direction of the twisting, one can now dynamically twist the ten spin-half $\psi^m$ variables to spin-zero $\G^m$ variables and spin-one $\Gb^m$ variables defined by
\eqn\defgam{\G^m =  {1\over{2(\l\lb)} }\g (\l\g^m\g^n\lb) \psi_n, \quad
\Gb^m =  {1\over{2(\l\lb)}} {1\over \g} (\lb\g^m\g^n\l) \psi_n,}
so that 
\eqn\defpsi{\psi^m = \g \Gb^m + {1\over\g} {{(\l\g^m\g^n\lb)}\over{2(\l\lb)}} \G_n.}
$\bar\G^m$ will be constrained to satisfy
\eqn\Gcons{\Gb^m (\g_m\lb)^\a =0,} 
and
since $\psi^m$ of \defpsi\ is invariant under the gauge transformation $\d\G_m = \e\g_m\lb$
generated by \Gcons, 
only half of the $\G^m$ and $\Gb^m$ components are
independent.
After performing this twisting and expressing $\psi^m$ in terms of $\G^m$ and $\Gb^m$,
$GSO(+)$ states only depend on even powers of the $\g$ ghost.
So it will be useful to define 
\eqn\ghdef{\gh \equiv (\g)^2}
which carries conformal weight $-1$, and define $\bh$ of conformal weight $+2$ to
be the conjugate momentum to $\gh$. 

In terms of $(\G^m, \Gb_m, \gh, \bh)$, the worldsheet action and stress tensor are
\eqn\actthr{S = \int d^2 z (\half\p x^m\bar\p x_m + \Gb^m \bar\p\G_m + \bh \bar\p\gh + b\bar\p c }
$$+ p_\a \bar\p\t^\a +  v^m \bar\p u_m + w'_\a \bar\p \l^\a + \wb^{'\a} \bar\p \lb_\a + s^\a \bar\p r_\a ),$$
\eqn\stressthr{T = -\half \p x^m \p x_m -\Gb^m \p \G_m - \bh\p\gh - \p(\bh \gh) -
b\p c - \p (bc) }
$$ - p_\a \p\t^\a -  v^m \p u_m - w'_\a \p \l^\a - \wb^{'\a} \p \lb_\a - s^\a \p r_\a, $$
where 
\eqn\defbh{\bh = {1\over {2 \g}} \b + {1\over {2\g^2}}\G^m \Gb_m,}
$$w'_\a = w_\a - {1\over {4\g^2 (\l\lb)}} \G^m \G^n [ (\lb\g_{mn})_\a +  {{(\l\g_{mn}\lb)}\over{(\l\lb)}} \lb_\a ],$$
$$\wb^{'\a} = \wb^\a - {{\g^2}\over {4 (\l\lb)}} \Gb^m \Gb^n (\l\g_{mn})^\a +
{1\over{2(\l\lb)}} \G^m\Gb^n (\l\g_m\g_n)^\a,$$
and the conjugate momenta $\bh$, $w'_\a$ and $\wb^{'\a}$ of \defbh\
have been defined to have no singular OPE's with 
$\G^m$ and $\Gb_m$.
Note that the twisting of the $\psi^m$ variables to $(\G^m, \Gb_m)$ variables
shifts their central charge contribution from
$+5$ to $-10$, and is compensated by the replacement of the $(\b,\g)$ with $(\bh,\gh)$ variables which shifts their central charge contribution from $+11$ to $+26$.

Expressing $Q$ of \brstfinal\ in terms of the twisted variables of \defgam\ and \ghdef, 
one obtains
\eqn\brstf{ Q = 
 \int dz (\l^\a d_\a +{\wtb}^{'\a} r_\a + {{(\l\g^m\g^n\lb)}\over{2(\l\lb)}}\Pi_m\G_n + {{\G_m \G_n}\over{4(\l\lb)}} [(\lb\g^{mn}\p\t)
 +{{(\l\g^{mn}\lb)}\over{(\l\lb)}}(\lb\p\t)] + c T - b c \p c}
 $$+ \gh [b+ \Gb^m \Pi_m + {{\Gb_m \Gb_n}\over{4(\l\lb)}} (\l\g^{mn} r) + s^\a \p\lb_\a +
 {{(\lb \g^{mn}\p\t)}\over{4(\l\lb)}} N'_{mn} + {{(\lb \p\t)}\over{4(\l\lb)}} (J' + 4u_n v^n) + (\l\g^m\p\t) v_m ]$$
 $$+ u_m [ \Gb^m +{1\over{2(\l\lb)}}\lb\g^m d - {{(\lb\g^{m} \g^{np} r)}\over{8(\l\lb)^2}} (N'_{np} +{1\over{\gh}}\G_n\G_p) ])$$
 where $N'_{mn} = \half (\l\g_{mn} w')$, $J' = - \l^\a w'_\a$, and 
 \eqn\defwtb{\wtb^{' \a} \equiv \wb^{' \a} + {1\over{2(\l\lb)}} (\l\g_m\g_n)^\a (u_m v_n - \G_m\Gb_n).}
 Since $\wtb^{' \a}$ of \defwtb\  commutes with the constraints 
 $v^m (\g_m\lb)^\a = 0$ and $ \Gb^m (\g_m\lb)^\a=0$ of \constv\ and \Gcons\ 
 up to the gauge transformation
 $\d \wtb^{' \a} = f^m (\lb\g_m)^\a$, one can easily verify
 that $Q$ of \brstf\ also commutes with these constraints.
  
 The BRST operator of \brstf\ is closely related to the simplified form of the composite
 pure spinor $b$ ghost found in \BerkovitsPLA. The third line of \brstf\ is $u_m$ times
 the constraint in \BerkovitsPLA\ for $\Gb^m$, and the second line of \brstf\ contains
 $\gh$ times the composite $b$ ghost expressed in terms of $\Gb^m$. 
After applying the similarity transformation 
${\cal O} \to e^{-S} e^{-R} {\cal O} e^{R} e^S$ where
\eqn\simthree{R = \int dz  {1\over{2(\l\lb)}}
\G^m [\lb\g^m d - {{(\lb\g^{m} \g^{np} r)}\over{4(\l\lb)}} (N'_{np} +{1\over{\gh}}\G_n\G_p)],}
$$ S = - \int dz \gh v^m (\Pi_m + {{(\l\g^m\g^n r)}\over{4(\l\lb)}}\Gb_n),$$
\brstf\ reduces to
\eqn\brstg{ Q = 
 \int dz (\l^\a d_\a +\wtb^{'\a} r_\a + \gh [b-B + v_m(\l\g^m)_\a \p ({{\G_n (\g^n\lb)^\a}\over{2(\l\lb)}} )] + u_m \Gb^m + c T - b c \p c)}
 where $B$ is the usual composite pure spinor $b$ ghost (ignoring normal ordering terms)
 \eqn\compb{B = 
 - s^\a\p\lb_\a  + {{\lb_\a (2
\Pi^m (\g_m d)^\a-  N'_{mn}(\g^{mn}\p\t)^\a
- J' \p\t^\a)}\over{4(\lb\l)}} }
$$- {{(\lb\g^{mnp} r)(d\g_{mnp} d +24 N'_{mn}\Pi_p)}\over{192(\lb\l)^2}}
+ {{(r\g_{mnp} r)(\lb\g^m d)N^{' np}}\over{16(\lb\l)^3}} -
{{(r\g_{mnp} r)(\lb\g^{pqr} r) N^{' mn} N'_{qr}}\over{128(\lb\l)^4}}.$$

Finally, the similarity transformation 
${\cal O} \to e^{-U} {\cal O} e^{U} $ where
\eqn\simfour{U = \int dz c (B - v_m (\l\g^m)_\a \p ({{\G_n (\g^n\lb)^\a}\over{2(\l\lb)}})  - \bh\p c)}
transforms \brstg\ into
\eqn\brsth{ Q = 
 \int dz (\l^\a d_\a +\wtb^{'\a} r_\a + \gh b+ u_m \Gb^m).}
 Note that this last similarity transformation shifts the Virasoro $b$ ghost to
 \eqn\shiftb{ e^{-U} b e^U = b + B - v_m (\l\g^m)_\a \p ({{\G_n (\g^n\lb)^\a}\over{2(\l\lb)}})  - \bh\p c - \p (\bh c).}
 The usual quartet argument implies that the cohomology of \brsth\ is independent
 of $(u_m, v^m)$, $(\G_m, \Gb^m)$, $(\gh, \bh)$, $(b,c)$, $(\lb_\a, \wtb^{'\a})$, and
 $(r_\a, s^\a)$, so one recovers the original pure spinor BRST operator
 $Q_{pure} =\int dz  \l^\a d_\a$.
 
 So after adding non-minimal spinors and twisting the spin-half $\psi^m$ variables into spin-zero and spin-one variables, the RNS BRST operator has been
 covariantly mapped into the pure spinor BRST operator. In the next two sections, this
 covariant map will be used to relate vertex operators and scattering amplitudes in the two formalisms.
 
 \newsec{Vertex Operators}  

In this section, massless vertex operators in the RNS and pure spinor formalisms will be related to each other using the covariant map of the previous section. After adding the
non-minimal spinor variables of \nrnsa\ to the RNS formalism, both Neveu-Schwarz and Ramond vertex operators can be constructed in the ``zero picture'' without spin fields or bosonization.
In this non-minimal RNS formalism, vertex operators in the $GSO(+)$ sector can be expressed in d=10
superspace and there is no difficulty with describing Ramond-Ramond backgrounds.
After dynamically twisting and gauge-fixing, these massless vertex operators
in the non-minimal RNS formalism reduce to the massless vertex operators in
the pure spinor formalism.

\subsec{Non-minimal RNS vertex operators}

After adding the non-minimal spinor variables of \nrnsa\ and performing the
similarity transformation of \simtwo, the non-minimal RNS BRST operator of \rnse\ takes the manifestly spacetime supersymmetric form 
\eqn\rnsz{Q = \int dz (\L^\a d_\a +{1\over {2 \g}} (\L\g^m\L)\psi_m + c T +   \g \psi^m \Pi_m + \g^2 (b+\O_\a\p\t^\a) - b c \p c ).}
To construct massless open superstring vertex operators in the ghost-number one cohomology of $Q$, the first step will
be to use the ``minimal coupling'' construction of Siegel \siegelclassical\ in which the
operators $[d_\a, \Pi_m, \p\t^\a]$ in $Q$ are replaced by the d=10 super-Yang-Mills superfields
$[A_\a(x,\t), - A_m(x,\t), - W^\a(x,\t)]$. These superfields satisfy the on-shell constraints \SiegelT\WittenT
\eqn\onshell{D_\a A_\b + D_\b A_\a = \g^m_{\a\b} A_m, \quad
D_\a A_m - \p_m A_\a = (\g_m)_{\a\b} W^\b, }
$$D_\a W^\b = \half (\g^{mn})_\a{}^\b \p_m A_n = {1\over 4}(\g^{mn})_\a{}^\b F_{mn}, $$
and are defined up to the gauge transformation
\eqn\gaugeg{\d A_\a = D_\a \Sigma, \quad \d A_m = \p_m \Sigma,}
where $D_\a = {\p\over{\p\t^\a}} + \half (\g^m\t)_\a \p_m$ is the d=10 supersymmetric
derivative. In components, one can gauge 
\eqn\comp{A_\a =\half (\g^m\t)_\a a_m + {1\over 3}(\g^m\t)_\a (\g_m\t)_\b \xi^\b + ..., \quad
A_m = a_m + (\g_m\t)_\a \xi^\a + ... ,}
$$W^\a = \xi^\a -\half (\g^{mn}\t)^\a \p_m a_n + ...,\quad
F_{mn} = \p_m a_n - \p_n a_m  + ...,$$
where $a_m(x)$ and $\xi^\a(x)$ are the onshell gluon and gluino fields satisfying
$\p^m \p_{[m} a_{n]} =0$ and $\p_m (\g^m\xi)_\a =0$, and $...$ 
denotes terms higher-order in $\t^\a$ which can be expressed in terms
of derivatives of $a_m$ and $\xi^\a$.

So the ``minimal coupling'' construction of Siegel predicts the massless vertex operator
\eqn\minimal{V_{min} = \L^\a A_\a (x,\t) - \g\psi^m A_m(x,\t) - \g^2 \O_\a W^\a (x,\t)}
$$+ c (\p\t^\a A_\a(x,\t) + \Pi^m A_m (x,\t) + d_\a W^\a (x,\t) ).$$
Using the constraints of \onshell, one finds that $Q V_{min}$ is nonzero and satisfies
\eqn\qvmin{Q V_{min} = Q [ \half c (\psi^m\psi^n - \half \L \g^{mn}\O) F_{mn}(x,\t) +
c \g\psi^m \O_\a \p_m W^\a(x,\t) ].}
So the minimal coupling construction needs to be slightly corrected and the
ghost-number one BRST-invariant vertex operator is 
\eqn\app{V = \L^\a A_\a (x,\t) - \g\psi^m A_m(x,\t) -\g^2 \O_\a W^\a (x,\t)} 
$$+ c (\p\t^\a A_\a(x,\t) + \Pi^m A_m (x,\t) + d_\a W^\a (x,\t) )$$
$$-\half c (\psi^m\psi^n - \half \L \g^{mn}\O) F_{mn}(x,\t) -
c \g\psi^m \O_\a \p_m W^\a(x,\t) .$$
The integrated BRST-invariant vertex operator of ghost-number zero is defined in the usual manner as $\int dz U \equiv \{\int dz b, V\}$, so the integrated vertex operator is 
\eqn\intu{\int dz U = \int dz [ \p\t^\a A_\a(x,\t) + \Pi^m A_m (x,\t) + d_\a W^\a (x,\t)}
$$+\half
 (-\psi^m\psi^n + \half \L \g^{mn}\O) F_{mn}(x,\t) -
 \g\psi^m \O_\a \p_m W^\a(x,\t)].$$
The term $\half(-\psi^m\psi^n + \half \L \g^{mn}\O) F_{mn}(x,\t) $ in \intu\ is expected
since when $F_{mn}$ is constant, the integrated vertex operator should be the Lorentz
generator 
\eqn\Lorentz{\int dz [ x^{[m} \p x^{n]} -\psi^m \psi^n -  \half (\t\g^{mn} p) +\half( \L\g^{mn}\O)]}
where the $x^{[m} \p x^{n]} - \half (\t\g^{mn} p)$ contribution to \Lorentz\ comes from the 
$\p\t^\a A_\a + \Pi^m A_m + d_\a W^\a$ terms in \intu. However, the presence of
the $- \g\psi^m \O_\a \p_m W^\a(x,\t) $ term in \intu\ is surprising and it would be useful to get a better understanding of this term.

By adding the integrated vertex operator of \intu\ to the open superstring worldsheet action of
\nrnsa, one can describe super-Yang-Mills backgrounds with both Neveu-Schwarz
and Ramond background fields turned on. And by taking the ``left-right product''
of two open superstring vertex operators and adding to the closed superstring worldsheet
action, one can describe d=10 supergravity backgrounds in the non-minimal
RNS formalism which include Ramond-Ramond background fields.

Despite the ${1\over \g}$ dependence in the similarity transformation of \simtwo\
and in the BRST operator of \rnse, the massless super-Yang-Mills vertex operator
of \app\ has no $1\over\g$ dependence and is therefore in the ``small'' Hilbert space.
And since all massive vertex operators in the $GSO(+)$ sector can be obtained from
OPE's of the super-Yang-Mills vertex operators, all physical vertex operators in the
$GSO(+)$ sector of the non-minimal RNS formalism can be constructed in the ``small''
Hilbert space.

On the other hand, the physical vertex operator for the Neveu-Schwarz tachyon
in the non-minimal RNS formalism has ${1\over \g}$ dependence and is 
\eqn\tach{V = e^{-R'} [(\g  +i c \psi^m k_m)e^{ik^m x_m} ] e^{R'} =
(\g  +i c (\psi^m - {1\over {2\g}}\L\g^m\t) k_m )e^{ik^m x_m}}
where $R'$ is defined in \simtwo. So it appears that vertex operators in the $GSO(-)$
sector of the non-minimal RNS formalism cannot be constructed in the ``small''
Hilbert space.

\subsec{Relation with RNS and pure spinor vertex operators}

To relate the non-minimal vertex operator of \app\ with the minimal RNS vertex operator
for the gluon, one gauges $\t^\a =\L^\a=0$ in \app\ to obtain
\eqn\rnsmin{V = -\g \psi^m a_m (x) + c (\p x^m a_m (x) - \psi^m\psi^n \p_m a_n (x))}
which is the standard RNS gluon vertex operator. However, because there are no spin fields in the non-minimal RNS vertex operators, it is unclear how to relate
the non-minimal and minimal RNS vertex operators for the gluino. 

To relate the non-minimal vertex operator of \app\ with the super-Yang-Mills vertex operator
in the pure spinor formalism, one gauges $u_m = \G_m = c = \gh =0$. In this gauge,
$\L^\a$ reduces to $\l^\a$ and the vertex operators of \app\ and \intu\ reduce to
\eqn\puremin{V = \l^\a A_\a (x,\t)}
$$\int dz U = \int dz [\p\t^\a A_\a(x,\t) + \Pi^m A_m (x,\t) + d_\a W^\a (x,\t)+
 {1\over 4} (\l \g^{mn} w) F_{mn}(x,\t)]$$
which are the standard unintegrated and integrated super-Yang-Mills vertex operators in the pure spinor formalism.

So the vertex operator of \app\ in the non-minimal RNS formalism 
provides a bridge between the usual RNS and pure spinor vertex operators. Surprisingly, the 
non-minimal vertex operators for Ramond states do not require spin fields or
bosonization, and the non-minimal vertex operators for states in the $GSO(-)$ sector
cannot be expressed in the ``small'' Hilbert space. 
It would be very useful to understand how to relate these non-minimal RNS vertex operators with the usual RNS vertex operators for 
Ramond states and $GSO(-)$ states.

\newsec{Scattering Amplitudes}

In this section, dynamical twisting will be argued to transform the
RNS amplitude prescription into the pure spinor amplitude prescription. So assuming
that dynamical twisting is an allowable field redefinition at the quantum level, the RNS
and pure spinor amplitude prescriptions are expected to be equivalent. However, it should
be stressed that there are various subtleties with both the RNS and pure spinor
amplitude prescriptions and the argument sketched here does not address these subtleties.
For example, the non-split nature of super-moduli space in the RNS formalism \Donagi\ makes
it difficult to compute multillop amplitudes using picture-changing operators. And in the
pure spinor formalism, the presence in multiloop 
amplitudes of poles when $(\l\lb)\to 0$ requires
special regulators  \BerkovitsNek\ which complicate the
computation of higher-genus terms that are not protected by supersymmetry.

In string theories with chiral bosons, functional integration over the chiral boson zero modes needs to be regularized. As long as the regularization method preserves BRST invariance, on-shell amplitudes are expected to be independent of the regularization method.
A convenient BRST-invariant method for regularizing the functional integration over chiral bosons is to insert a picture-changing operator for each chiral boson zero mode. Dynamical twisting modifies the structure of the chiral bosons and therefore modifies  the picture-changing operators used to regularize their functional integration. By taking
into account this modification coming from dynamical twisting, the RNS amplitude prescription can be related to the pure spinor amplitude prescription.

\subsec{Non-minimal RNS amplitude prescription}

In the RNS formalism, the $(\b,\g)$ chiral bosons carry conformal weight $({3\over 2}, -\half)$
and therefore have $(0,2)$ zero modes on a genus zero surface, $(1,1)$ zero modes
on a genus one surface, and $(2g-2,0)$ zero modes on a genus $g$ surface for $g>1$.
For each $\g$ zero mode, one needs to insert a ``picture-lowering'' operator \FriedanGE
\eqn\pY{Y_\g = c \d'(\g) = c \p \xi e^{-2\phi}}
where $\g = \eta e^\phi$ and $\b = \p\xi e^{-\phi}$. And for each $\b$ zero mode, one
needs to insert a ``picture-raising'' operator
\eqn\pZ{Z_\b = : \d(\b) Q(\b) : = \d(\b) (\p x^m \psi_m + ...).}

In the non-minimal RNS formalism, one also has the $(\L^\a, \O_\a)$ chiral bosons
of conformal weight $(0,1)$ which have $(16, 16g)$ zero modes on a genus $g$ surface.
Using the non-minimal BRST operator of \rnsz, one needs to insert for each $\L^\a$
zero mode the BRST-invariant picture-lowering operator
\eqn\pYa{Y_{\L^\a} = e^{-R} \d(\L^\a) \t^\a e^R = \d(\L^\a)\t^\a - \d'(\L^\a) c \t^\a \p\t^\a}
where $R$ is defined in \simone. And one needs to insert for each $\O_\a$ zero mode
the BRST-invariant picture-raising operator
\eqn\pZa{Z_{\O_\a} = : \d(\O_\a) [Q, \O_\a ] : = \d(\O_\a) (d_\a + ...).}

For the scattering of external gluons, one can verify that the picture-lowering
and picture-raising operators of \pYa\ and \pZa\ absorb all the zero modes of
$(\L^\a, \O_\a)$ and $(\t^\a, p_\a)$. Furthermore, the functional integral over the
bosonic non-zero modes of 
$(\L^\a, \O_\a)$ cancels the functional integral over the fermionic
non-zero modes of $(\t^\a, p_\a)$. So after performing the functional integration over the 
$(\L^\a, \O_\a)$ and $(\t^\a, p_\a)$ variables, the amplitude prescription for external gluon
scattering coincides with the usual RNS amplitude prescription. 

However, for scattering involving external gluinos, the non-minimal RNS prescription
is very different from the usual RNS prescription since the non-minimal Ramond
vertex operators do not contain spin fields or half-integer picture. It would be fascinating
to find a proof that scattering amplitudes involving external gluinos coincide in the 
non-minimal and minimal RNS formalisms. 

\subsec{ Pure spinor amplitude prescription}

To relate the non-minimal RNS formalism with the pure spinor formalism, one needs to 
dynamically twist the spin-half fermions $\psi^m$ into spin-zero and spin-one fermions
$\G_m$ and $\Gb^m$ using pure spinors $(\l^\a, \lb_\a)$ constructed from the
unconstrained spinor $\L^\a = \l^\a + {1\over{2(\l\lb)}} u_m (\g^m\lb)^\a$ of \defL. In addition,
one needs to replace the $(\b, \g)$ ghosts of conformal weight $({3\over 2}, -\half)$
with $(\bh, \gh)$ ghosts of conformal weight $(2,-1)$ where $\gh \equiv (\g)^2$.
As will now be argued, the different zero mode structure of chiral bosons created
by this dynamical twisting will modify the RNS scattering amplitude prescription into
the pure spinor scattering amplitude prescription.
So if dynamical twisting can be proven at the quantum level to be a consistent field
redefinition, the scattering amplitude prescriptions in the two formalisms should be
equivalent.

After dynamical twisting, the chiral bosons include the pure spinor variables
$(\l^\a, w_\a)$ and $(\lb_\a, \wb^\a)$ of conformal weight $(0,1)$, the $(u^m, v_m)$
variables of conformal weight $(0,1)$, and the $(\bh, \gh)$ variables of conformal
weight $(2, -1)$. Functional integration over the zero modes of the pure spinors
$(\l^\a, w_\a)$ and $(\lb_\a, \wb^\a)$ can be performed using the standard pure
spinor regulator \BerkovitsBT\
\eqn\regps{{\cal N} = e^{-\{Q, \t^\a \lb_\a + \sum_{I=1}^g w_{\a I} s_I^\a \} }
= e^{-(\l^\a\lb_\a + \t^\a r_\a + \sum_{I=1}^g (w_{\a I} \wb^\a_I + d_{\a I} s_I^\a) + ...)} }
where $(w_{\a I}, \wb_I^\a, s_I^\a, d_{\a I})$ for $I=1$ to $g$ are the
$g$ holomorphic zero modes of 
 $(w_{\a}, \wb^\a, s^\a, d_{\a})$ and 
 \eqn\brstnp{Q= \int dz (\l^\a d_\a + \wtb^{' \a} r_\a + u_m \Gb^m + \gh (b-B + ...) + cT - bc \p c )}
 is the BRST operator of \brstg.
For each zero mode of $u_m$, one needs to insert the picture-lowering operator
\eqn\pYc{Y_{u^m} = \d(u_m) \G_m.}
And for each zero mode of $v_m$, one needs to insert the picture-raising operator
\eqn\pZc{Z_{v_m} = :\d(v_m) [Q, v_m]: = \d(v_m)  \Gb_m .}
Finally, for each zero mode of $\gh$, one needs to insert
the picture-lowering operator 
\eqn\pYd{Y_{\gh} = \d(\gh) c.}
And for each zero mode of $\bh$, one needs to insert the picture-raising operator
\eqn\pZd{Z_{\bh} = :\d(\bh) [Q, \bh]: = \d(\bh)  (b - B + ...)}
where $B$ is the pure spinor $b$ ghost of \compb.

Since $(\bh, \gh)$ have the same conformal weight as
the $(b, c)$ Virasoro ghosts, they have the same number
of zero modes on the worldsheet. To reproduce the
pure spinor amplitude prescription, the picture-lowering
operators $Y_{\gh}$ of \pYd\ should be inserted on
the unintegrated vertex operators and the 
picture-raising operators $Z_\bh$ of \pZd\ should be inserted
on the $(3g-3)$ $b$ ghosts contracted with the Beltrami
differentials. With this choice, the contribution from
each unintegrated vertex operator is $c \d(\gh) (\l^\a A_\a (x,\t) + ...)$ and the contribution from each of the $(3g-3)$
Beltrami differentials is $b \d(\bh) (b - B + ...)$. 

After inserting all the picture-lowering and picture-raising operators of \pYc\ -- \pZd, functional integration over the $(u_m, v^m)$ variables cancels
the functional integration over the $(\G_m, \Gb^m)$ variables and functional integration over the
$(b,c)$ variables cancels the functional integration over the $(\bh, \gh)$ variables. The functional integral over the
remaining variables with the regulator of \regps\ reproduces the usual pure spinor
amplitude prescription where the $(3g-3)$ Beltrami
differentials are contracted with the $B$ operator of
\compb. 

So after dynamically twisting and inserting the appropriate picture-lowering and picture-raising operators to regularize the functional integration over the chiral boson zero modes, the non-minimal RNS scattering amplitude prescription reduces to the usual pure spinor amplitude prescription. 

But as was mentioned earlier, there are several
subtleties which have been ignored in this argument.
For example, the functional integral in the pure spinor
formalism at higher genus is singular if there are poles
of order $(\l\lb)^{-11}$ coming from the $(3g-3)$ pure
spinor $B$ ghosts \BerkovitsNek. And in the RNS formalism, the non-split structure of higher genus supermoduli space \Donagi\ complicates the computation using picture-changing operators. It would be very interesting if
these multiloop subtleties in the two formalisms could
be related to each other using the covariant map of this paper.

\vskip 10pt
{\bf Acknowledgements:}
I would like to thank
Nikita Nekrasov and Edward Witten for useful discussions, Sebastian Guttenberg for
informing me about reference \gutt, 
and 
CNPq grant 300256/94-9
and FAPESP grants 09/50639-2 and 11/11973-4 for partial financial support.

\listrefs
\end